# Ising Superconductivity and Quantum Phase Transition in Macro-Size Monolayer NbSe$_2$

Ying Xing,[†,‖,Δ] Kun Zhao,[‡,#,Δ] Pujia Shan,[†,#] Feipeng Zheng,[†,#] Yangwei Zhang,[†,#]

Hailong Fu,[†,#] Yi Liu,[†,#] Mingliang Tian,[§] Chuanying Xi,[§] Haiwen Liu,[⊥] Ji Feng,[†,#]

Xi Lin,[†,#] Shuaihua Ji,[‡,#,*] Xi Chen,[‡,#] Qi-Kun Xue,[‡,#] Jian Wang[†,‡,#,*]

[†]*International Center for Quantum Materials, School of Physics, Peking University, Beijing 100871, China*

[‡]*State Key Laboratory of Low-Dimensional Quantum Physics, Department of Physics, Tsinghua University, Beijing 100084, China*

[§]*High Magnetic Field Laboratory, Chinese Academy of Sciences, Hefei 230031, China*

[‖]*Beijing Key Laboratory of Optical Detection Technology for Oil and Gas, China University of Petroleum, Beijing 102249, China*

[⊥]*Department of Physics, Beijing Normal University, Beijing 100875, China*

[#]*Collaborative Innovation Center of Quantum Matter, Beijing 100084, China*

ABSTRACT

Two-dimensional (2D) transition metal dichalcogenides (TMDs) have a range of unique physics properties and could be used in the development of electronics, photonics, spintronics and quantum computing devices. The mechanical exfoliation technique of micro-size TMD flakes has attracted particular interest due to its simplicity and cost effectiveness. However, for most applications, large area and high quality films are preferred. Furthermore, when the thickness of crystalline films is down to the 2D limit (monolayer), exotic properties can be expected due to the quantum confinement and symmetry breaking. In this paper, we have successfully prepared macro-size atomically flat monolayer $NbSe_2$ films on bilayer graphene terminated surface of 6H-SiC(0001) substrates by molecular beam epitaxy (MBE) method. The films exhibit an onset superconducting critical transition temperature ($T_c^{onset}$) above 6 K, 2 times higher than that of mechanical exfoliated $NbSe_2$ flakes. Simultaneously, the transport measurements at high magnetic fields reveal that the parallel characteristic field $B_{c//}$ is at least 4.5 times higher than the paramagnetic limiting field, consistent with Zeeman-protected Ising superconductivity mechanism. Besides, by ultralow temperature electrical transport measurements, the monolayer $NbSe_2$ film shows the signature of quantum Griffiths singularity when approaching the zero-temperature quantum critical point.

TEXT

Quasi-2D superconductors such as ultrathin films with thickness down to monolayer [1-7] and composites interfaces[8-9], have remained an active topic in recent years due to fundamental research interests and potential applications. Nevertheless, just a few monolayer crystalline superconductors can be prepared successfully on special substrates since fluctuations can destroy the long range correlation of superconductivity in 2D systems. Recently, TMDs as natural layered materials have provided a new platform to study superconductivity due to the tunable nature of the superconducting properties coexistent with other collective electronic excitations, as well as strong intrinsic spin-orbit coupling. The bulk crystals of TMDs are formed of monolayers bound to each other by Van-der-Waals attraction, which makes it feasible to experimentally study monolayer TMDs.

$2H-NbSe_2$, one kind of TMDs, is found to be superconducting even in its freestanding monolayer.[10-13] More interestingly, Zeeman-protected Ising superconductivity is expected in monolayer $NbSe_2$ due to the non-centrosymmetric structure with in-plane inversion symmetry breaking and strong spin−orbit coupling. Very recently, Ising superconductivity with the anomalous large in-plane critical magnetic field has become one important direction in crystalline 2D superconductors[7]. In recent experiments on exfoliated micron-size $NbSe_2$ monolayers[10, 11, 13], the coexistence of charge density wave (CDW) and the superconducting phase was observed down to the monolayer limit but the $T_c$ of monolayer $NbSe_2$ got significantly suppressed (less than 3.1 K) compared with its bulk value (7.2 K).

Superconductor-insulator (metal) transition (SIT/SMT), a paradigm of quantum phase transition, is an important topic in condensed matter physics.[14-20] Despite of efforts over last few decades, there still remain many open issues, such as different critical exponents signifying different universality classes and various values of critical points found in different materials.[21] Recent observations of the quantum Griffiths singularity in thin Ga films[22] and LaAlO$_3$/SrTiO$_3$(110) (LAO/STO) interface[23] shed a new light on SIT/SMT[21] and have turned to one important topic in 2D superconductors[7]. Verifying quantum Griffiths singularity in Ising superconductors would not only demonstrate the universal property of quantum Griffiths singularity, but also help to understand the underlying mechanism of the 2D superconductors where Ising superconductivity and quantum Griffiths singularity coexist.

In this paper we successfully prepared atomically flat monolayer NbSe$_2$ films ($\sim$ 0.6 nm thick) on bilayer graphene terminated surface of 6H-SiC (0001) substrates by MBE method. The high quality films are uniform and macro-size large (mm$^2$). Here, monolayer means one Se-Nb-Se sequence and each unit cell of NbSe$_2$ consists of two Se-Nb-Se sequences. By low temperature and high magnetic field electrical transport measurements, the Ising superconductivity is found in NbSe$_2$ monolayers, with $T_c^{\text{onset}}$ up to 6.61 K and $B_{c//}$ ($T = 0$) up to 37.22 T. The observed superconductivity anisotropy and Berezinski-Kosterlitz-Thouless (BKT)-like transition reveal the 2D nature of NbSe$_2$ films. Through systematic ultralow temperature transport measurements, the monolayer NbSe$_2$ films manifest magnetic field induced SMT and the critical exponent of SMT diverges as $T$ approaching zero, indicating quantum Griffiths singularity.

Figure 1(a) shows the typical atomic-resolved scanning tunneling microscope (STM) image (18 nm × 18 nm) on NbSe$_2$ monolayer, from which a 3 × 3 CDW superlattice can be clearly seen at 80 mK. Large-scale (1.9 μm × 1.9 μm) STM topographic image with atomically flat surface and regular steps originating from the SiC (0001) surface is shown in Figure 1(b). The flat terraces are monolayer 2H-NbSe$_2$ and the small islands are two monolayers NbSe$_2$ with random orientation (not 2H phase). The monolayer NbSe$_2$, bilayer graphene and SiC (0001) substrate are all Van der Waals coupling between each other. Due to this weak coupling, the monolayer Se-Nb-Se is more like freestanding state. Before being transferred out of MBE high vacuum chamber, an amorphous Se capping layer with the thickness of 20 nm was deposited on the monolayer $NbSe_2$ at 80 K to protect the film from degrading in ambient atmosphere. These formed $Se/NbSe_2$/bilayer graphene/SiC heterostructure.

Figure 2(a) exhibits temperature dependence of sheet resistance ($R_s$) of sample 1 at zero magnetic field. With the temperature decreasing, the $R_s$ of monolayer $NbSe_2$ rapidly increases at 120 K, saturates around 50 K, then begins to decrease at 6.61 K ($T_c^{onset}$), and drops to zero within the instrumental resolution at 2.40 K ($T_c^{zero}$) (Figure 2(b)). The inset of Figure 2(a) schematically depicts the diagram for transport measurements. The similar $R_s(T)$ property for sample 2 is displayed in Figure S1(a) with $T_c^{onset}$ up to 6.95 K. The Se/bilayer graphene/SiC heterostructure under the growth conditions identical to $Se/NbSe_2$/bilayer graphene/SiC heterostructure exhibits insulating behavior (Figure S1(b)). This confirms that the superconductivity only comes from the monolayer $NbSe_2$. A magnetic field up to 15.50 T perpendicular to the film was applied (Figure 2(b)) and magnetoresistance isotherms (perpendicular magnetic field) were measured at temperature from 0.50 K to 10.00 K (Figure 2(c), full data shown in Figure S2(a)). It is clearly evident that the increasing magnetic field gradually destroys the superconductivity. When the magnetic field is parallel to the film (see Figure 2(d)), $B_c(T)$ is apparently much higher than that in perpendicular field. 15 T can hardly destroy the superconductivity of monolayer $NbSe_2$. To obtain $B_c$ at lower temperatures, we measured the magnetoresistance of the same sample by applying steady high magnetic field up to 35 T. For $T = 0.35$ K, 35 T high field still cannot destroy the superconductivity of the monolayer $NbSe_2$ completely (Figure 2(e)). We define the characteristic field $B_c(T)$ at a given temperature at the 80% of normal resistance for both perpendicular and parallel magnetic fields, and summarize in Figure 2(f). With $T/T_c$ down to 0.2, $B_c(T)$ can be fitted by $B_{c\perp}(T) \propto 1 - T/T_c$ and $B_{c//}(T) \propto (1 - T/T_c)^{1/2}$ respectively,[24, 25] which yield $B_{c\perp}(0)$ = 2.87 T and $B_{c//}(0)$ = 37.22 T. The anisotropic parameter $\varepsilon = B_{c//}(0)/B_{c\perp}(0)$ is about 13 for NbSe$_2$ monolayer but only 3.2 for bulk NbSe$_2$[26]. BKT-like transition has also been observed (Figure S3). Such observations confirm the 2D nature of the superconductivity in monolayer NbSe$_2$.

Theoretically, we investigate the electron-phonon coupling of freestanding monolayer NbSe$_2$ without charge density wave modulation by density-functional theory and density-functional perturbation theory calculations within local density approximation. The critical temperature $T_c$ calculated using Mcmillan formula[27] falls in range between 3.8 and 4.1 K, which is close to our experimental result ($T_c^{onset} \sim$ 6.61 K and $T_c^{zero} \sim$ 2.40 K for sample 1). We also qualitatively investigated the influence of bilayer graphene on the electronic band structure of the monolayer NbSe$_2$ and found that the graphene has little influence on the electronic band structure of the NbSe$_2$. (Figure S4)

In the 2D limit regime, the orbital effect is restricted in parallel magnetic field. The $B_c$ is only determined by Pauli-limiting field $B_p$, which originates from the Zeeman splitting. In general, the Pauli-limiting field is strong enough to break Cooper pairs and destroy superconductivity. $B_p$, is written as $B_p = g^{-1/2}\Delta/\mu_B$, where $g$ is the Landé $g$-factor, $\mu_B$ is the Bohr magneton, and $\Delta$ is the superconducting gap. For BCS superconductors, $B_p$ can be simply rewritten as $B_p = 1.84 T_c$ when assuming a $g$-factor equals to 2. For our MBE grown monolayer NbSe$_2$ films, $B_p$ is estimated to be 6.35 T ($T_c$ (80%) $\sim$ 3.45 K) by using $g \sim 2$ in previous reports[10]. If we use g factor value of bulk NbSe$_2$ ($\sim$ 1.2)[28], then $B_p$ should be 8.21 T. Surprisingly, as revealed in $B_{c2}$-$T$ phase diagram in the inset of Figure 2(f), the monolayer NbSe$_2$ could withstand an applied magnetic field as strong as 37.22 T ($T = 0$). Thus, $B_{c//}(0)$ is at least 4.5 times of $B_{\mathrm{p}}$ ($g \sim$ 1.2). This phenomenon has been reported in gated $MoS_2$ flakes[29, 30] and mechanical exfoliation fabricated micro-size $NbSe_2$ flakes[10]. In non-centrosymmetric monolayer $NbSe_2$ with considerable spin-orbit interactions, the spin-orbit interaction split the spin states and manifest as a strong effective internal magnetic fields. The spin-orbit interaction experienced by a moving electron with momentum $k$ is proportional to $k \times E \cdot \sigma$, where $E$ is the electric field experienced by the electron and $\sigma$ denoted the Pauli matrices. In monolayer $NbSe_2$, the in-plane inversion symmetry is broken and the electrons can experience effective in-plane electric field. Hence, the spins of the pairing electrons are strongly locked to the out-of-plane orientation by an effective Zeeman field. Instead of damaging superconductivity, this special type of internal magnetic field due to strong spin-orbit interaction is able to protect the superconducting electron pairs under high external magnetic fields. This kind of superconductor is called "Ising superconductor".

We note that the high critical magnetic field in low temperature regime has been observed in heavy Fermion superconductors[31] and organic superconductors[32]. The mechanism was interpreted as Fulde-Ferrell-Larkin-Ovchinnikov (FFLO) state[33]. For the FFLO phase to appear, the compounds must have a Maki parameter[34] $\alpha = \sqrt{2}\, B_{\mathrm{orb}}/B_{\mathrm{p}}$ larger than 1.8 such that the upper critical field can easily approach the Pauli paramagnetic limit $B_{\mathrm{p}}$. Simultaneously, the system must be in the clean limit $\xi << l$, since the FFLO state is readily destroyed by impurities.[31] Calculations show that in anisotropic superconductors, the FFLO state might lead to an enhancement of the upper critical field between 1.5 and 2.5 times of the Pauli paramagnetic limit field[32, 35]. In perpendicular magnetic field cases, the characteristic field $B_{c\perp}(0)$ of $NbSe_2$ film is estimated to be 2.87 T, smaller than the Pauli paramagnetic limit field (8.21 T). Besides, the Maki parameter $\alpha_\perp = 0.44$ is smaller than 1.8. In parallel magnetic field cases, the characteristic field $B_{c//}(0) \sim 37.22$ T is at least 4.5 times of Pauli paramagnetic limit field, which exceeds the theoretical predictions of 1.5 ~ 2.5 times[32, 35]. Therefore, the chance of the existence of FFLO state in monolayer NbSe$_2$ is little.

The sample 3 (with $T_c^{\text{onset}} \sim 6$ K, Figure S5(a)) was measured in ultralow temperature environment by Physical Property Measurement System (Figure S5, down to 2.00 K) and Dilution Refrigerator (Figure 3, down to 0.025 K). Figure S5(b) depicts the $R_s(T)$ curves at various magnetic fields. With the increasing field, the superconducting film gradually tunes to a metal. The magnetoresistance isothermals seem to cross a point from 2.00 K to 4.50 K (Figure S5(c)). However, systematic ultralow temperature data display a series of magnetoresistance crossing points between 0.025 K and 1.30 K, which form a continuous line of SMT "critical" points (Figure 3(b)). We summarize the SMT "critical" points $B_c^{'}$ in Figure 3(c). The black squares are crossing points of $R_s(B)$ curves at every two adjacent temperatures. The $R_s$ plateaus extracted from $R_s(T)$ curves in Figure 3(a) are shown as red dots, at which the $\mathrm{d}R_s/\mathrm{d}T$ changes sign for a given magnetic field. The blue line is the linear fitting curve. As we can see, the $B_c^{'}$ slightly diverges the linear tendency in ultralow temperature region ($T < 0.264$ K). The linear extrapolation gives $B_c^{'}(0 \text{ K}) = 2.940$ T, obviously below the actual experimental data ($> 3.200$ T).

For SIT, the sample critical resistance on phase transition is the quantum resistance for pairs ($h/4e^2 \sim 6450$ Ω) and the critical exponent remains a constant[36]. This is not in agreement with our experimental results. The critical resistance of monolayer NbSe$_2$ film is much smaller than $h/4e^2$, indicating the unpaired normal electrons also contribute to the conductance. Such unpaired electrons can originate from the dissipation effect, which gives rise to the SMT with critical resistance much smaller than $h/4e^2$ [37]. Subsequent theoretical investigations reveal that the quenched disorder dramatically changes the scaling behavior of SMT[38], and result in activated scaling identical to that of the random transverse field Ising model, in which the dynamical exponent $z$ continuously varies when approaching the quantum critical point[39]. This active scaling behavior can be regarded as the "quantum version" of Griffiths singularity, which is called quantum Griffiths singularity[40]. In previous reports on SITs and SMTs, the magnetic resistance isotherms cross in one or two points, which are considered as quantum criticality.[14-20] However, in recent work on 2D superconducting Ga films[22], the similar continuous line of SMT "critical" points and the dynamical critical exponent divergence were detected, which experimentally revealed quantum Griffiths singularity of SMTs in Ga films. [7, 21, 22]

We then analyze our data referring to the finite size analysis method[22] (See supplemental information for details). The resulting $B$ dependence of effective "critical" exponent $zv$ is summarized in Figure 3(d). In relatively high temperature regime, $zv$ increases slowly with magnetic field. In the ultralow temperature regime, $zv$ grows quickly and diverges when the critical magnetic field ($B_c^*$) is approaching. We fit the experimental values ($zv > 1$) as a function of $B$ by the activated scaling law $zv = C|B_c^* - B|^{-v\psi}$ [41], where the correlation length exponent $v \approx 1.2$, the tunneling critical exponent in two dimensions $\psi \approx 0.5$ and $C$ is a constant. The activated scaling behavior can fit the experimental data very well (Figure 3(d)). This indicates the existence of infinite randomness QPT in monolayer NbSe$_2$, consistent with the quantum Griffiths singularity[41-46] behavior. Thus, for the first time, quantum Griffiths singularity is detected in superconductors at 2D limit, i.e. monolayer NbSe$_2$ films.

In summary, several innovations are displayed in this paper. (i) The high-quality macro-size atomically flat monolayer $NbSe_2$ films are successfully grown on bilayer graphene/SiC by MBE method and the film exhibits $T_c^{onset}$ above 6 K by electronic transport study, higher than reported values[10, 11, 13] on mechanically exfoliated $NbSe_2$ monolayers. (ii) By the measurements at high magnetic fields up to 35 T, we undoubtedly reveal the superconducting survivability under a strong parallel magnetic field with $B_{c//}(0) > 4.5$ $B_p$ as $T/T_c$ down to 0.2 and demonstrate Ising superconductivity in macro-size monolayer $NbSe_2$. (iii) The magnetic field driven SMT is detected in monolayer $NbSe_2$ films and the quantum phase transition exhibits the signature of quantum Griffiths singularity. The coexistence of Ising superconductivity and quantum Griffiths singularity at 2D limit might become an important topic in future for further understanding 2D superconductivity. Besides the innovations mentioned above, monolayer $NbSe_2$ has also been theoretically proposed to be a new platform to create topological superconductivity and Majorana Fermions[47-49]. Our findings will definitely stimulate more investigations on 2D TMD superconductors.

FIGURES

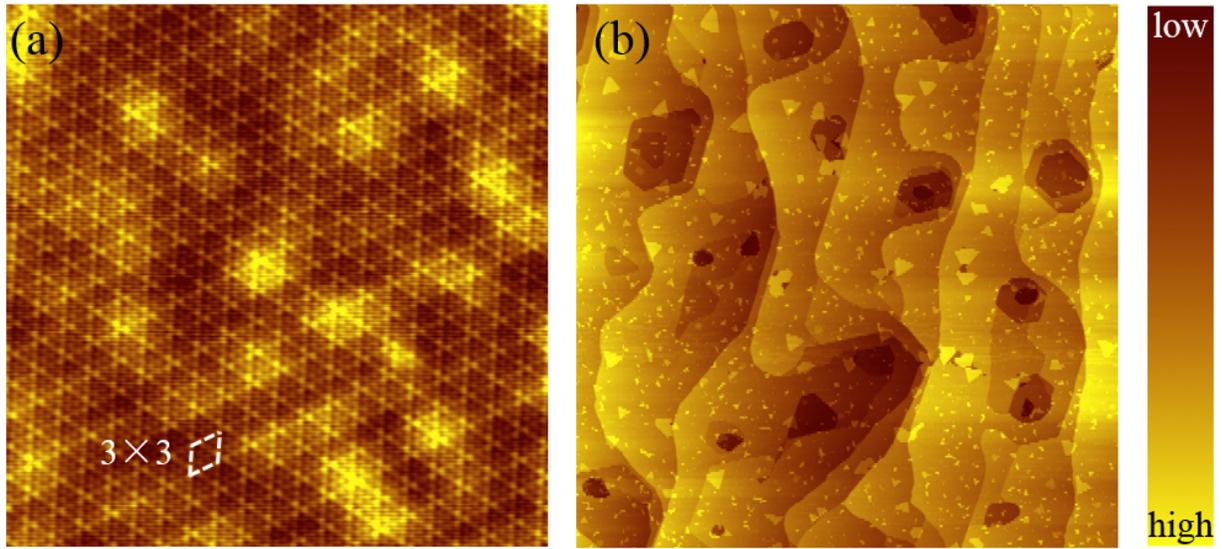

**Figure 1.** (a) The atomic-resolved STM image on monolayer NbSe$_2$ film (18 nm × 18 nm, sample bias $V_s$ = 10 mV, tunneling current $I_t$ = 100 pA, temperature $T$ = 80 mK). The 3×3 CDW superlattice is fully and uniformly developed for monolayer NbSe$_2$. (b) The typical topography of monolayer NbSe$_2$ on graphene/SiC(0001) (1.9 μm × 1.9 μm, $V_s$ = 3.0 V, $I_t$ = 20 pA, $T$ = 300 K).

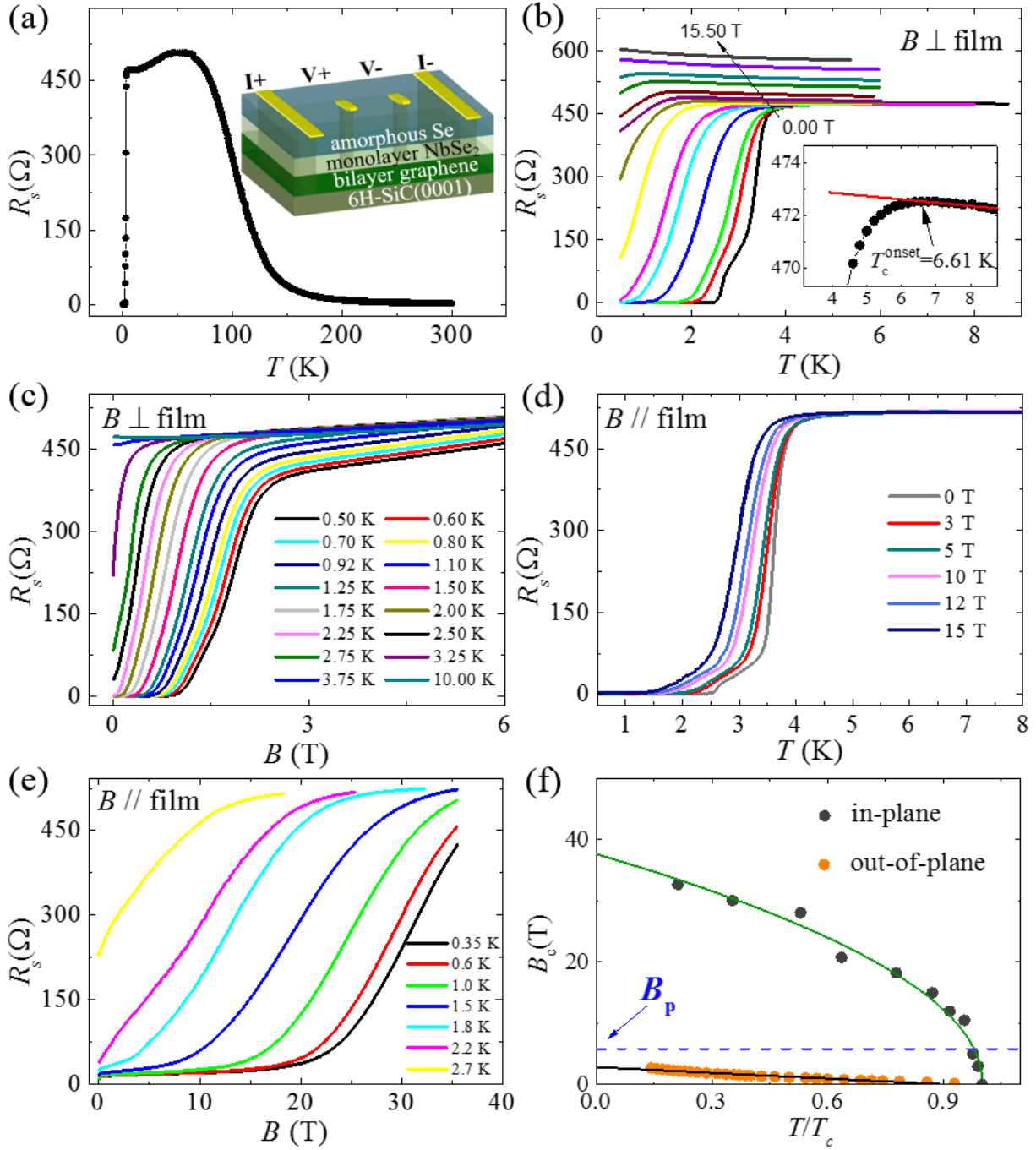

**Figure 2.** Electrical transport measurements of monolayer NbSe₂ film. (a) Temperature dependence of the sheet resistance of the heterostructure Se/NbSe₂/BL Graphene/6H-SiC at zero magnetic field. The inset depicts the schematic diagram for heterostructure and standard four-

probe configuration for electrical measurements. (b) $R_s(T)$ curves for various perpendicular magnetic fields up to 15.5 T (the magnetic fields from bottom to top are: 0.00 T, 0.10 T 0.20 T, 0.50 T, 0.80 T, 1.00 T, 1.50 T, 2.00 T, 3.00 T, 5.00 T, 8.00 T, 10.00 T, 13.00 T, 15.50 T.). The $R_s$(T) curve at zero magnetic field reveals $T_c^{zero}$ = 2.40 K. The inset is the zoom-in image of $R_s(T)$ curve from 4 K to 8 K, showing $T_c^{onset}$= 6.61 K. (c) Magnetoresistance measured in perpendicular magnetic field at selected temperatures from 0.5 K to 10 K. (d) $R_s(T)$ and (e) $R_s(B)$ characteristics at parallel magnetic fields (the magnetic field is parallel to the film up to 35 T). (f)The characteristic magnetic field $B_c$ for both perpendicular and parallel field. $B_{c\perp}(0)$ = 2.87 T, $B_{c//}(0)$ = 37.22 T. $B_{c\perp}(T)$ and $B_{c//}(T)$ are extracted from (c) and (e) as the fields at which the sample resistance drops to 80% of the normal state resistance. The green line is the fitting curve using $B_{c//}(T) \propto (1\text{-}T/T_c)^{1/2}$. The black line is linear fitting by $B_{c\perp}(T) \propto 1\text{-}T/T_c$. The dashed blue line indicates the Pauli limit field $B_p$ = 6.35 T.

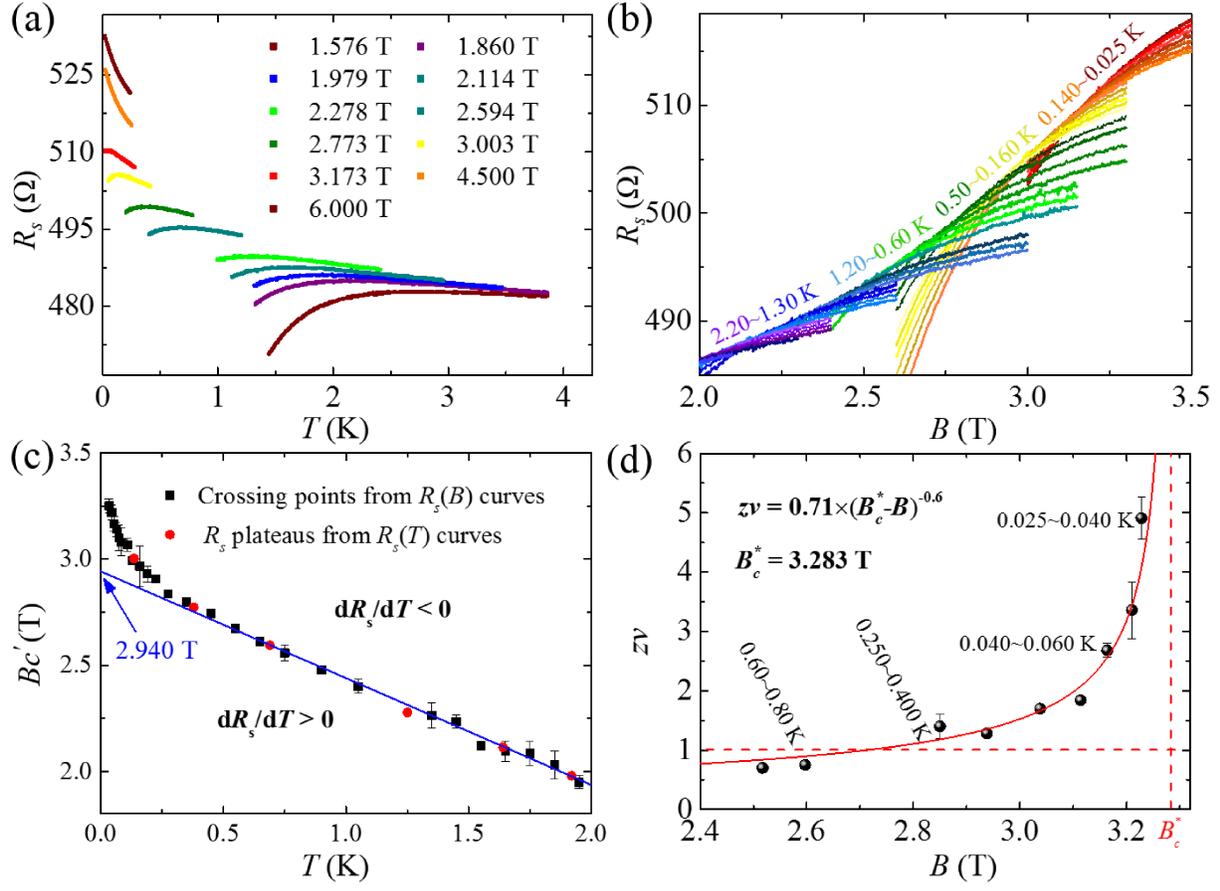

**Figure 3.** (a) $R_s(T)$ curves at different magnetic fields. The temperatures at $R_s$ maximum($dR_s/dT$ = 0) in each curves are: 0.136 K, 0.380 K, 0.69 K, 1.25 K, 1.64 K, 1.92 K, 2.18 K, and 2.72 K, respectively. (b) The magnetic field dependence of $R_s$ at various temperatures from 0.020 K to 2.20 K. (c) The critical magnetic field $B_c^{'}$ at various temperatures, extracted from (a) and (b). The black squares are crossing points of $R_s(B)$ data at adjacent temperatures in (b); the red dots are the $R_s$ maximums on $R_s(T)$ curves in (a). The crossing points are well consistent with $R_s$ maximums. The blue line is the linear fitting curve. (d) SMT exponent $zv$ as a function of magnetic field. The red curve is a fitting based on the activated scaling law equation shown in (d).

ASSOCIATED CONTENT

**Supporting Information**.

Sample preparation, Transport measurements and BKT-like transition, Theoretical calculations, Finite size scaling analysis, Two-parameter scaling fitting procedure for activated scaling law (word)

AUTHOR INFORMATION

**Corresponding Authors**

*E-mail: jianwangphysics@pku.edu.cn.

*E-mail: shji@mail.tsinghua.edu.cn.

Author Contributions

$^{\triangle}$Y. X. and K. Z. contributed equally to this work.

Notes

The authors declare no competing financial interest.

ACKNOWLEDGMENT

We thank Dingping Li for useful discussions. This work was supported by the National Basic Research Program of China (Grants No. 2013CB934600, the Research Fund for the Doctoral Program of Higher Education (RFDP) of China (20130001110003), the Open Research Fund

Program of the State Key Laboratory of Low-Dimensional Quantum Physics, Tsinghua University under Grant No. KF201501, the Open Project Program of the Pulsed High Magnetic Field Facility (Grant No. PHMFF2015002), Huazhong University of Science and Technology, and the Science Foundation of China University of Petroleum, Beijing (2462017YJRC012).

# Supplemental Information

## Ising Superconductivity and Quantum Phase Transition in Macro-Size Monolayer NbSe$_2$

Ying Xing,[†,∥,Δ] Kun Zhao,[‡,#,Δ] Pujia Shan,[†,#] Feipeng Zheng,[†,#] Yangwei Zhang,[†,#] Hailong Fu,[†,#]

Yi Liu,[†,#] Mingliang Tian,[§] Chuanying Xi,[§] Haiwen Liu,[⊥] Ji Feng,[†,#] Xi Lin,[†,#] Shuaihua Ji,[‡,#,*]

Xi Chen,[‡,#] Qi-Kun Xue,[‡,#] Jian Wang[†,‡,#,*]

[†]*International Center for Quantum Materials, School of Physics, Peking University, Beijing 100871, China*

[‡]*State Key Laboratory of Low-Dimensional Quantum Physics, Department of Physics, Tsinghua University, Beijing 100084, China*

[§]*High Magnetic Field Laboratory, Chinese Academy of Sciences, Hefei 230031, China*

[∥]*Beijing Key Laboratory of Optical Detection Technology for Oil and Gas, China University of Petroleum, Beijing 102249, China*

[⊥]*Department of Physics, Beijing Normal University, Beijing 100875, China*

[#]*Collaborative Innovation Center of Quantum Matter, Beijing 100084, China*

[Δ]Ying Xing and Kun Zhao contributed equally to this work.

*Corresponding authors: jianwangphysics@pku.edu.cn; shji@mail.tsinghua.edu.cn

## Sample preparation

The growth of sample and characterization of morphology were performed in a molecular beam epitaxy and scanning tunneling microscopy combined UHV system (Unisoku). Monolayer NbSe$_2$ films were grown on the top of bi-layer graphene with Boron-doped 6H-SiC(0001) substrates. The bilayer graphene was prepared by the thermal desorption approach in UHV condition. Monolayer NbSe$_2$ film was grown by co-evaporating high-purity Nb and Se from an electron-beam evaporator and a standard Knudsen cell, respectively. The growth was set under a Se-rich condition and monitored by *in-situ* reflection high-energy electron diffraction (RHEED). The ion flux of Nb source was 10 nA, giving rise to a slow growth rate of 2.5 MLs per hour.

## Transport measurements and Berezinski – Kosterlitz – Thouless (BKT) - like transition

For *ex situ* transport measurements, the contacts on the NbSe$_2$ film through the protection layer can be achieved by pressing the indium on the top surface. Unless specifically mentioned, the electrical resistance and magnetoresistance were acquired using a four-point geometry with excitation current limited to 5 μA. The NbSe$_2$ films are about 2 mm wide and 6 mm long, and the current electrodes were thin strips across the width of the film to assure a uniform current distribution over the width of the sample. The resistance and magnetoresistance were carried out in a Physical Property Measurement System (PPMS) with a Helium-3 cryostat down to 0.5 K and up to 16 T. For varying the angle between the sample plane and the external magnetic field, the devices were mounted on a rotation holder. Ultralow temperature measurements were done in a Dilution Refrigerator MNK126-450 system, (a.c. technique, 17 Hz, base temperature < 0.010 K) with $T$ down to 0.020 K, magnetic field up to 14.000 T using 0.6 mT/s sweep rate, and 50 nA excitation current.

The $V(I)$ characteristics shown in Fig. S3(a) was measured at temperature ranging from 0.60 K to 8.00 K at zero magnetic field. At low temperatures, the $V(I)$ curves show a well-defined superconducting critical current $I_c$. At 0.60 K, $I_c$ is 0.53 mA, corresponding to critical current density $J_c \sim 4.2 \times 10^4$ A/cm$^2$ if proximity effect on graphene can be neglected. $J_c$ is in the same order with the bulk value[1]. If the proximity effect makes graphene superconducting, then the superconducting thickness layer should be 1.31 nm and the critical the current density is about $1.92 \times 10^4$ A/cm$^2$. These two values are comparable. In addition, the sample shows signatures of the BKT-like behavior[2, 3], such as a $V \propto I^\alpha$ power-law dependence. As revealed by Fig. S3(b), the exponent $\alpha$ approaches 3 at about 2.54 K, indicating $T_{BKT}$. This value is quite close to $T_c^{zero} \sim$ 2.40 K in $R_s(T)$ curves.

## Theoretical calculations

In our electron-phonon (e-ph) calculation, norm-conserving pseudopotential was used to model the interactions between valence electrons and ionic cores of both Nb and Se atoms. The Kohn-Sham valance states were expanded in the plane wave basis set with a kinetic energy truncated at 90 Ry. Spin-orbital coupling was taken into account. An over 20 Å vacuum range was inserted, to ensure sufficient separation between the periodic images of the monolayer NbSe$_2$. Firstly, converged electronic density of ground state and dynamical matrix were computed on a $\Gamma$-centered $\mathbf{k}$-grid of 18×18×1 for the sampling of electron Brillouin zone (BZ) and a $\Gamma$-centered $\mathbf{q}$-grid of 9×9×1 for the sampling of phonon BZ respectively using Quantum Espresso package[4]. Subsequently, the e-ph matrix elements were computed on the above $\mathbf{k}$- and $\mathbf{q}$-grids. Then the e-ph matrix was interpolated onto the much denser grids with both $\mathbf{k}$- and $\mathbf{q}$-grid of 200×200×1 through Wannier interpolation using EPW package[5, 6]. Next, the imaginary part of phonon self-energy and the mode-resolved e-ph coupling constants were computed on the above denser girds. Finally, the critical temperature $T_c$ was computed according to Mcmillan formula.

The crystal of the NbSe$_2$ used in the calculations was depicted in Fig. S4(a). Each Nb atom sits at the center of a triangular prism formed by six of its nearest Se atoms. The e-ph properties can be computed based on the isotropic approximation to Migdal-Eliashberg theory [7, 8]. In this framework, the phonon self-energy due to e-ph coupling can be expressed as

$$\Pi_{\mathbf{q}\nu} = \sum_{mn\mathbf{k}} w_{\mathbf{k}} |g_{mn}^{\nu}(\mathbf{k}, \mathbf{q})|^2 \frac{f(\epsilon_{n\mathbf{k}}) - f(\epsilon_{m\mathbf{k+q}})}{\epsilon_{n\mathbf{k}} - \epsilon_{m\mathbf{k+q}} - \omega_{\mathbf{q}\nu} + i\eta} \tag{1}$$

where $f(\epsilon_{n\mathbf{k}})$ is Fermi-Dirac distribution. $\epsilon_{n\mathbf{k}}$ is the electronic energy at a wave vector $\mathbf{k}$ with a band index n and $\omega_{\mathbf{q}\nu}$ is the phonon frequency at a wave vector $\mathbf{q}$ with a branch index $\nu$. $g_{mn}^{\nu}(\mathbf{k}, \mathbf{q}) = <m\mathbf{k+q}|\Delta V_{\mathbf{q}\nu}|n\mathbf{k}>$ is the e-ph coupling matrix element, where $\Delta V_{\mathbf{q}\nu}$ is the variation of the self-consistent potential due to the atomic displacements of a phonon mode with $\nu$ and $\mathbf{q}$. The phonon linewidth $\gamma_{\mathbf{q}\nu}$ is directly associated with the imaginary part of the phonon self-energy and the mode-resolved e-ph coupling constant $\lambda_{\mathbf{q}\nu}$ can be expressed in terms of $\gamma_{\mathbf{q}\nu}$ as $\lambda_{\mathbf{q}\nu} = \frac{1}{\pi N_F} \frac{\gamma_{\mathbf{q}\nu}}{\omega_{\mathbf{q}\nu}^2}$, where $N_F$ is the electronic density of states at Fermi level. The e-ph coupling constant $\lambda = \sum_{\mathbf{q}\nu} \lambda_{\mathbf{q}\nu}$. Finally, the critical temperature $T_c$ can be computed using Mcmillan formula [9]:

$$T_c = \frac{\omega_{\log}}{1.2} \exp\left[ \frac{-1.0 \; (1+\lambda)}{\lambda(1-0.6 \; \mu^*) - \mu^*} \right] \qquad (2)$$

where $\omega_{\log}$ is a characteristic phonon frequency, $\lambda$ is an electron-phonon coupling constant and $\mu^*$ represents an effective Coulomb potential (chosen between 0.1 and 0.15). The resultant values of $T_c$ fall in range between 3.8 and 4.1 K ($\mu^*$ chosen between 0.1 and 0.15), which is close to our experimental result ($T_c^{\text{onset}} \sim 6.61$ K and $T_c^{\text{zero}} \sim 2.40$ K).

We also qualitatively investigated the influence of bilayer graphene on the electronic bandstructure of the monolayer NbSe$_2$ [10]. The computed in-plane parameter of bilayer graphene and monolayer NbSe$_2$ are 2.446 and 3.398 Å respectively, indicating a serious lattice mismatch ($\sim 28\%$). To tackle this problem, we built up a superlattice of a 3×3×1 supercell of monolayer NbSe$_2$ on top of a 4×4×1 supercell of bilayer graphene, reducing the mismatch down to $\sim 4\%$. In this system, the in-plane parameter of the supercell of bilayer graphene was scaled to the one of the NbSe$_2$ and the superlattice was depicted in Fig. S4(b). The interlayer distance between the NbSe$_2$ and the graphene was obtained by minimizing the total energy of the superlattice. The optimized interlayer distance is 3.42 Å. Spin-orbit coupling was taken into account. The computed bandstructure of the superlattice (black lines) was depicted in Fig. S4(c). The components of the bands of the graphene near Fermi level (0 eV) were marked with thick and yellow lines. The bandstructure of the graphene without the NbSe$_2$ was depicted in the same pane with red lines. The Fermi levels of the two systems were aligned. Furthermore, the bandstructure of the NbSe$_2$ without the graphene was drawn in pane Fig. S4 (d) with red lines. The solid black lines in this pane are the same as the ones in pane (c). Compared the yellow lines and the red lines displayed in pane Fig. S4 (c), we find that the NbSe$_2$ is slightly electronically doped after taking the graphene into account. Despite of this, the graphene has little influence on the electronic bandstructure of the NbSe$_2$ as depicted in pane Fig. S4 (d).

## Finite size scaling analysis

The $R_s$ can be described by the phenomenological scaling law: $R(\delta,T) = R_c F(\delta T^{-1/\nu z})$, where $\delta = B - B_c^{'}$ is the deviation from the critical magnetic field, $R_c$ is the critical resistance, $F(\mu)$ is some universal function of $\mu$ with $F(0) = 1$, $\nu$ is the correlation length exponent, and $z$ is the dynamical critical exponent. Small critical region formed by three adjacent $R_s(B)$ curves can be treated approximately as one "critical" point ($B_c^{'}$, $R_c$). For instance, $B_c^{'}$ =3.228 T ($R_c$ =534.18 $\Omega$) is an approximate crossing point for transition region of three $R_s(B)$ curves of 0.025 K, 0.030 K and 0.040 K. We define $R(\delta,T_0) = R_c F(\delta)$ for the lowest temperature $T_0$, and rewrite the equation above as $R(\delta,T) = R_c F(\delta t) = R(\delta t,T_0)$, where the unknown parameter $t = (T/T_0)^{-1/z\nu}$ can be determined at each temperature by performing a rescale of the $\delta$ axis of resistance curves $R(\delta,T)$ to match the lowest $T_0$ curve $R(\delta,T_0)$. The effective "critical" exponent $z\nu$ for the temperature regime 0.025 K ~ 0.040 K is then determined to be 4.91 from the definition $t = (T/T_0)^{-1/z\nu}$. Nine representative crossing points on the $R_s(B)$ critical boundary are selected(Fig. S6), and the resulting magnetic field dependence of effective "critical" exponent $z\nu$ is summarized in Fig. 3(d).

## Two-parameter scaling fitting procedure for activated scaling law

In the quantum Griffiths phase, the activated scaling law results in a typical energy scale $\omega$, which obeys $\ln(\omega_0/\omega) \sim \xi^{\psi}$ with energy cut-off $\omega_0$. Here, $\xi \sim \delta^{-\nu}$ is the correlation length, $\delta = B - B_c$ is the deviation from the "critical" magnetic field $B_c$ and $\psi$, $\nu$ denote the tunneling exponent and correlation exponent, respectively. The $\rho_s(B)$ data follows the general two-parameter scaling law: $\rho_s(B,T) = \rho_c \Phi\left(\delta^{\nu\psi} \ln \dfrac{T^*}{T}\right)$, where $\rho_c$ is the critical resistivity and $\Phi$ is an arbitrary function[11]. $T^*$ is the characteristic temperature and set as 2.40 K($T_c^{\text{zero}}$). For the realistic experimental data, due to the finite size correction at finite temperature, the irrelevant scaling parameter can influence the crossing point of $\rho_s(B,T)$, thus the "critical" resistivity $\rho_c$ (crossing point for a bunch of $\rho_s(B,T)$ curves) manifests as a temperature dependent quantity as shown in Fig. S6. By performing a rescale of $\delta$ axis of $\rho_s$ to match the lowest $T_0$ curve in each group, $T$-dependent of $\nu\psi$ (around 0.6) can be then determined (Fig. S7). The fitting result is consistent with the original prediction of activated scaling law in reference [8], and also confirms the simplified version of activated scaling law used in Fig. 3(d) in the main text.

**Figures**

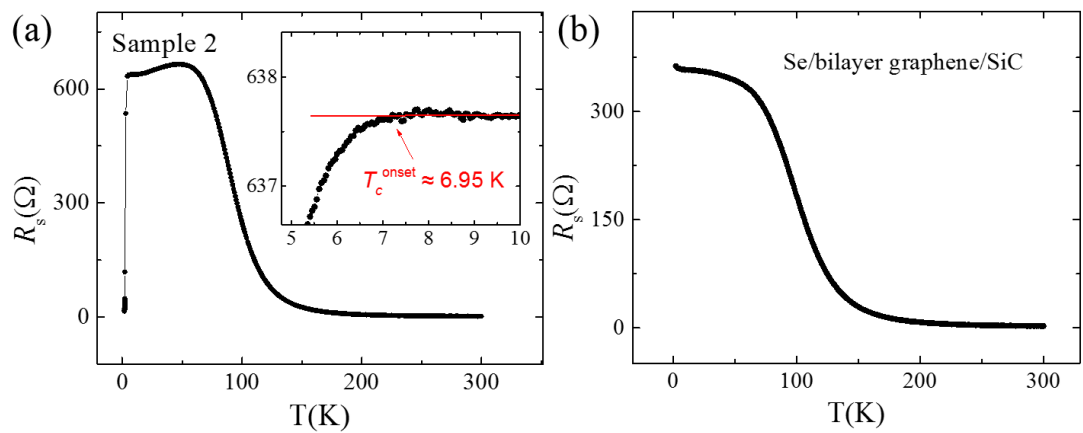

Figure S1 $R_s(T)$ at zero magnetic field of Se/NbSe$_2$/BL Graphene/6H-SiC (sample 2) (a) and Se/BL Graphene/6H-SiC (b), respectively.

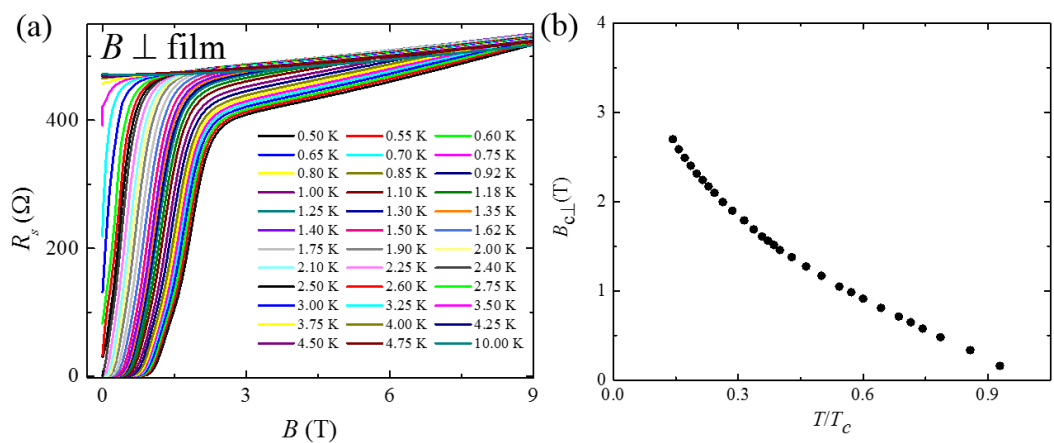

Figure S2 (a) Full data of $R_s(B)$ of sample 1 in the main text. (b) $B_{c\perp}(T)$ data in Fig.2(f), showing linear dependence in high temperature region and a slight upward deviation in low temperature region.

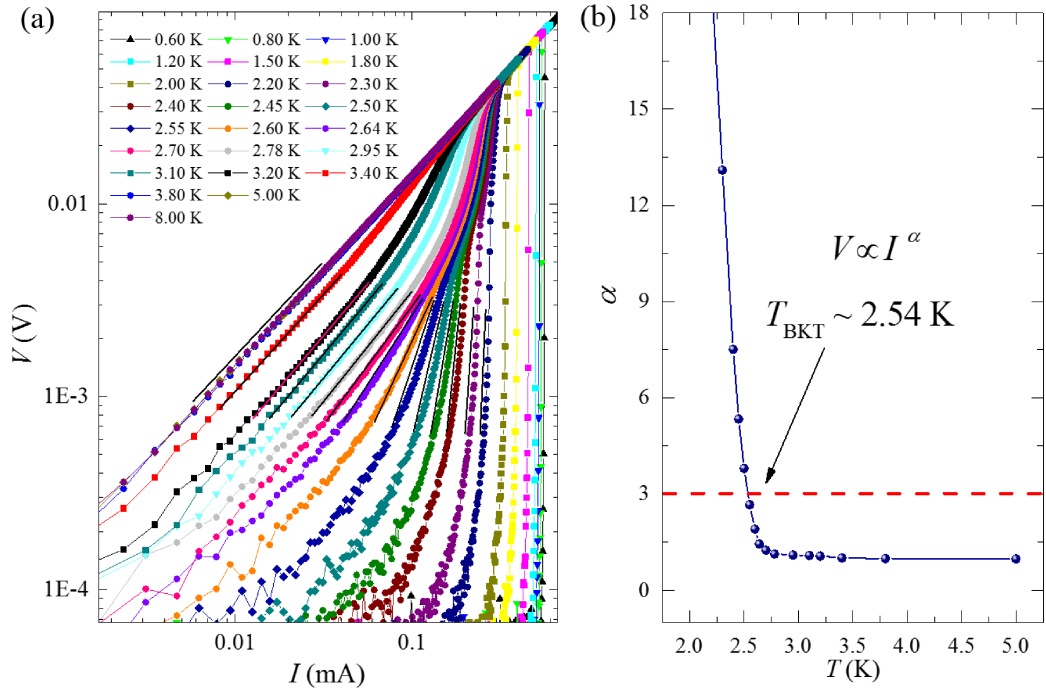

Figure S3 BKT-like transition. (a) $V(I)$ characteristics at various temperatures plotted on a logarithmic scale at $B = 0$ T. The solid lines are fits of the data in the transition. The two red solid lines correspond to $V \sim I$ and $V \sim I^3$ dependences and show that 2.50 K < $T_{BKT}$ < 2.55 K; (b) Temperature dependence of the power law exponent $\alpha$, as deduced from the fits show in (a).

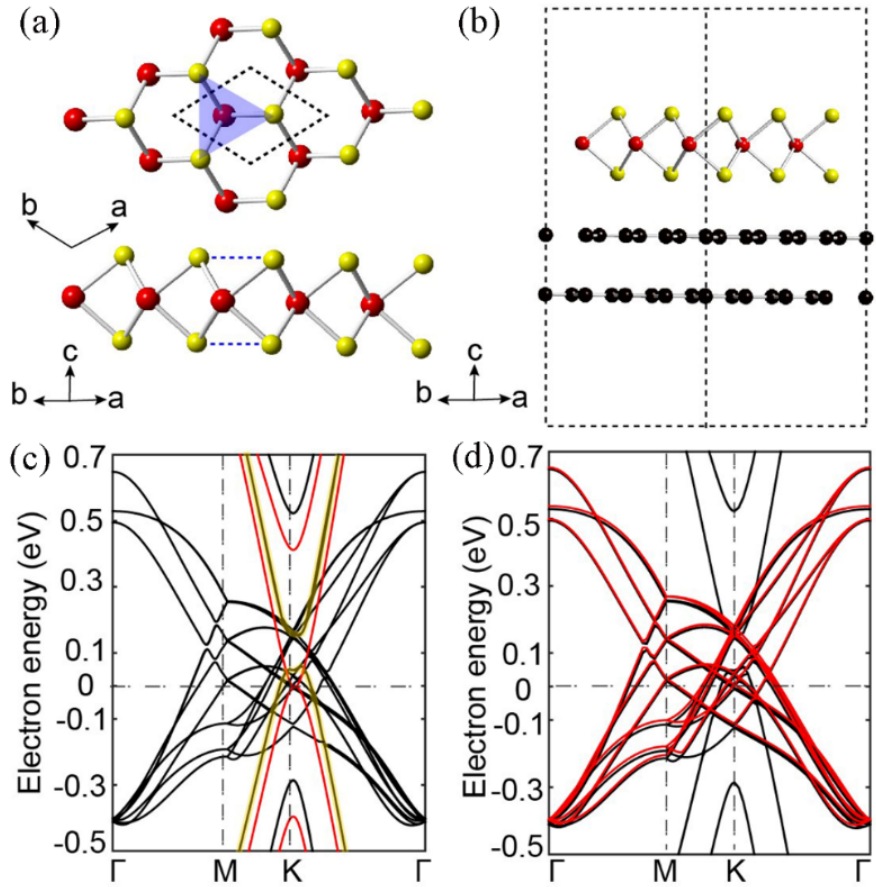

Figure S4 (a) Top and side view of the geometry structure of monolayer NbSe₂. The red and yellow balls stand for Nb and Se atoms respectively. The dashed black parallelogram represents a unit cell. The semitransparent trigon with blue colour represents the triangular prism around an Nb atom formed by six of its nearest Se atoms. The dashed blue lines bellow indicate the prism from side view; (b) Side view of the computational model of a superlattice, consisting of a 3×3×1 supercell of monolayer NbSe₂ on a 4×4×1 supercell of bilayer graphene. The black balls stand for carbon atoms. The dashed black lines represent a unit cell; Bandstructure of the superlattice shown in pane (b) was displayed in pane (c) (solid black lines). Among these bands, the components of the bilayer graphene near Fermi level (0 eV) were marked with two thick yellow lines. Red lines in pane (c) represent the bandstructure of the graphene without the NbSe₂. Fermi levels of the above two systems were aligned and marked with a dashed black line at 0 eV; The solid black lines in pane (d) are equal to the ones in pane (c) and the red lines in pane (d) represent the bandstructure of the NbSe₂ after removing the graphene. The Fermi levels of the two systems were also aligned.

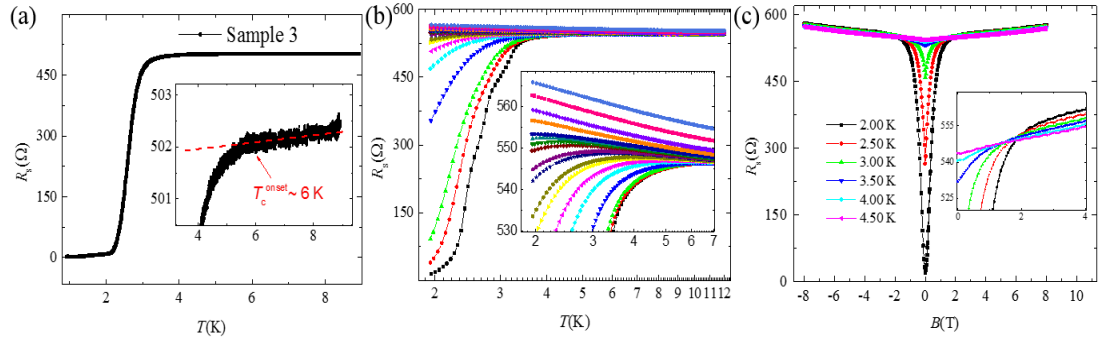

Figure S5 Transport data of sample 3 measured by Physical Property Measurement System. (a) $R_s(T)$ curves for sample 3; (b) $R_s(T)$ curves at different perpendicular magnetic fields of 0.00 T, 0.10 T, 0.20 T, 0.50 T, 0.80 T, 1.00 T, 1.20 T, 1.30 T, 1.50 T, 1.60 T, 1.80 T, 1.90 T, 2.00 T, 2.10 T, 2.50 T, 3.00 T, 5.00 T; (c) Magnetoresistance isotherms (2.00 K - 4.50 K) in perpendicular magnetic field. The insets of (a)(b)(c) are zoom in images.

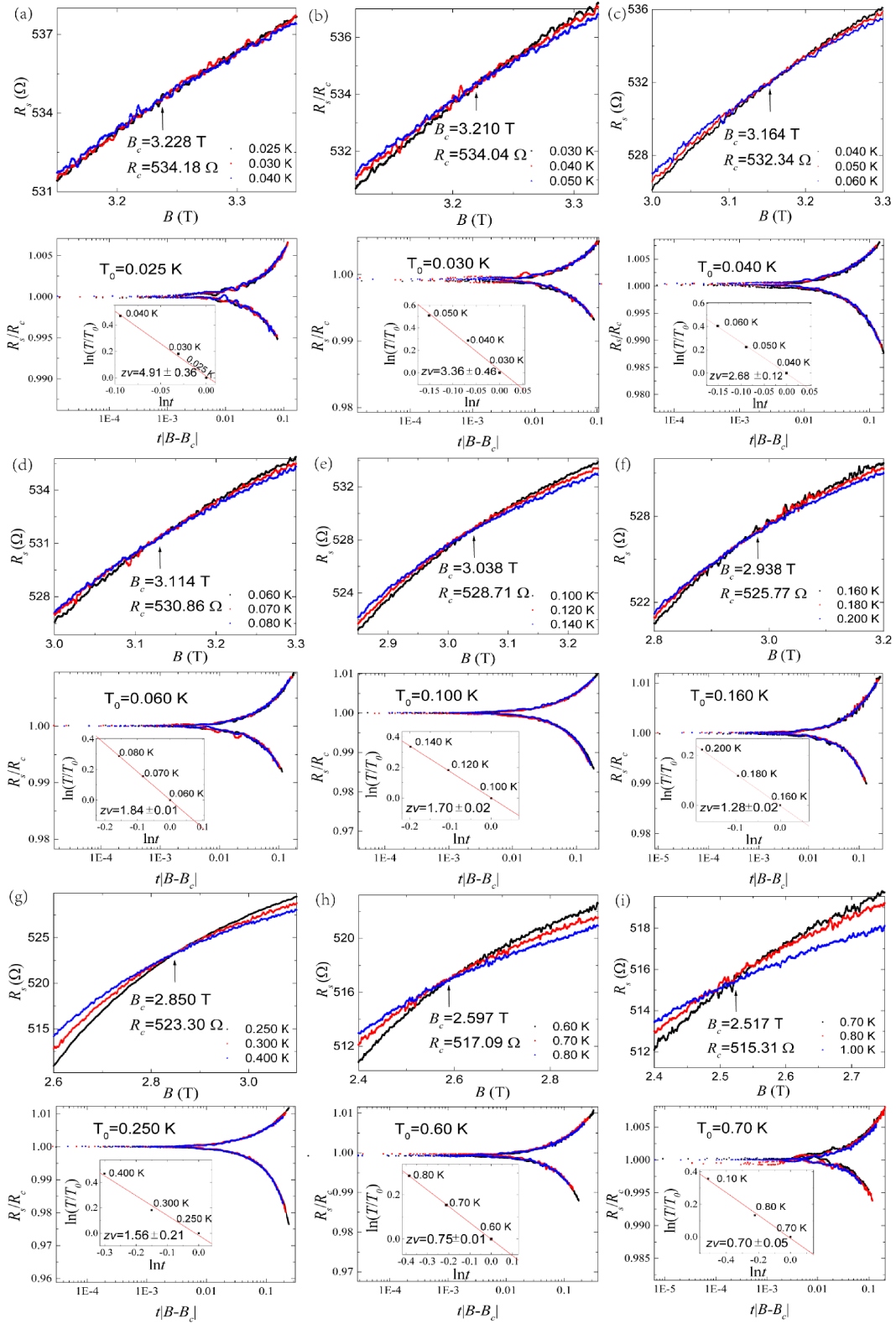

Figure S6 *zv* fitting (0.025 ~ 1.00 K). (a)(b)(c)(d)(e)(f)(g)(h)(i) The top columns reveal $R_s$ as a function of magnetic field at the superconductor-metal transition boundary at different temperatures. The bottom columns are normalized $R_s$ as a function of the scaling variable |B-$B_X$|$(T/T_0)^{-1/zv}$. The insets are the temperature dependence of the scaling parameter $t$ ($t = (T/T_0)^{-1/zv}$).

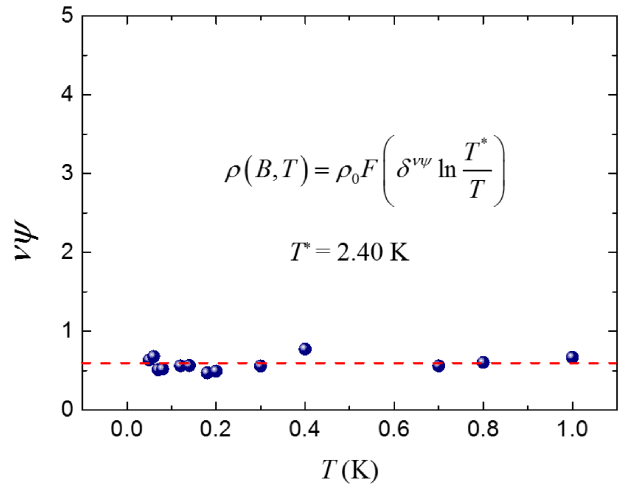

Figure S7 Temperature dependence of $v\psi$. The dashed red line indicates $v\psi = 0.6$.